\documentclass[12pt]{article}

\usepackage[cm]{fullpage}
\usepackage{amssymb}

\title{On the Post-linear Quadrupole-Quadrupole Metric}
\author{Francisco Frutos-Alfaro \\ 
\small{School of Physics and Space Research Center of the 
University of Costa Rica}
\\ Michael Soffel \\
\small{Technical University Dresden and Lohrmann Observatory}}
\date{\today}

\begin{document}
\maketitle

\abstract{
The Hartle-Thorne metric defines  a reliable spacetime for most
astrophysical purposes, for instance for the simulation of slowly rotating
stars. Solving the Einstein field equations, we added terms of second order
in the quadrupole moment to its post-linear version in order to compare it 
with solutions found by Blanchet in the frame of the multi-polar 
post-Minkowskian framework. We first derived the extended Hartle-Thorne metric 
in harmonic coordinates and then showed agreement with the corresponding 
post-linear metric from Blanchet.

We also found a coordinate transformation from the post-linear Erez-Rosen
metric to our extended Hartle-Thorne spacetime. It is well known that the
Hartle-Thorne solution can be smoothly matched with an interior perfect fluid
solution with physically appropriate properties. A comparison among these
solutions provides a validation of them. It is clear that in order to
represent realistic solutions of self-gravitating (axially symmetric) matter
distributions of perfect fluid, the quadrupole moment has to be included as a
physical parameter.
}



\section{Introduction}

\noindent
In 1968, Hartle and Thorne (HT) \cite{HT,Quevedo} proposed an approximate
solution to the Einstein field equations (EFE) intended to
represent the gravity field of neutron stars with mass, rotation and
quadrupole moment as parameters. Berti {\it et al.} \cite{Berti} compared the
HT metric with the Manko \cite{Mankoa,Mankob} (exact solutions) and
Cook-Shapiro-Teukolsky metrics \cite{CST} (numerical solution), and showed that
it is safe to use the HT metric, since it gives excellent results even for
the innermost stable circular orbits with fast spin periods. Moreover, the
exterior HT metric can be smoothly matched with a physically reasonable
interior one. This provides realistic models of stars and for this reason, it
is often used to validate exact and approximate solutions of EFE.

\noindent
Stationary exact solutions of EFE in the vacuum case are
characterized by two families of multipole-moments: mass- and spin-moments,
see for example \cite{HPM,Mankoa,Pachon,QM}. Some of these solutions might be
appropriate to represent stellar objects where the field moments can be related 
with corresponding body moments as integrals over the field generating sources.
Moreover, the post-linear approximation of these metrics must be
compatible with the HT solution. Quevedo {\it et al.} and Frutos {\it et al.} 
compared the HT solution with exact and approximate solutions of the EFE with a
quadrupole moment $ Q $ \cite{Frutos,Frutos2,Quevedo} of first order.
Comparisons with the second order in $ Q $ of these solutions are still
missing.

\noindent
Geroch and Hansen (GH) defined a procedure to find the field multipole moments 
of static and stationary spacetimes \cite{Geroch,Hansen}. Alternative 
definitions of relativistic multipole moments were given by Simon and Beig 
\cite{Simon} and by Thorne \cite{Thorne}. It is important to mention that the 
GH multipole moments \cite{Geroch,Hansen} are equivalent to the Thorne moments 
for stationary systems \cite{Guersel}.  Using the Ernst formalism
\cite{Carmeli,Ernst}, Fodor {\it et al.} found an elegant method to find the 
multipole moments of a given spacetime \cite{Fodor}. This method was later 
generalized by Hoenselaers and Perj\'es \cite{Hoenselaers}. The relevance of 
taking the correct relativistic multipole moments of numerical spacetimes
for modelling astrophysical objects such as neutron stars was discussed
by Pappas and Apostolatos \cite{Pappas}, who used a method due to 
Ryan  \cite{Ryan} to derive the multipole moments.

\noindent
Nowadays, the use of harmonic coordinates is costumary, since the
form of the transfomed metric tensor using harmonic coordinates has a special
structure that allows to read off the Thorne moments directly even for the
non-stationary case \cite{HPM,Thorne}.
The multi-polar post-Minkowskian formalism (MPM) that was developed
by Blanchet, Damour and Iyer \cite{BD,DI} is also formulated in harmonic 
coordinates. Applying this formalism Blanchet found spacetimes containing
mass-quadrupole and quadrupole-quadrupole terms \cite{Blanchet1,Blanchet2}.
The main goal of this paper is to compare these results with the ones we get
from a static HT approximation with squared quadrupole moment.

\noindent
The paper is organized as follows. In the second section, we
briefly describe the HT metric. We find a new expanded version of the HT metric 
with a squared quadrupole moment, in the third section. In the
fourth section, the harmonic coordinates for this HT solution are obtained
and the metric is expressed in these. In the fifth section, it is shown that
our HT harmonic metric and the Blanchet metric coincide at our level of
approximation. Finally,  a coordinate transformation is found from the
post-linear version of the Erez-Rosen metric
\cite{Carmeli,Doroshkevich,Winicour,Young,Zeldovich} to this HT solution with 
no rotation in the sixth section.

\section{The Hartle-Thorne Metric}

\noindent
The Hartle-Thorne metric \cite{HT,Quevedo} is an approximate solution of
vacuum EFE that describes the exterior of any slowly and
rigidly rotating, stationary and axially symmetric body. The metric is given
with accuracy up to the second order terms in the body's angular momentum,
and first order in its quadrupole moment. It therefore has three parameters:
mass $ M $, spin $ S $ and quadrupole-moment $ Q $. The HT solution is given by

\begin{equation} \label{ht} d {s}^2 = {g}_{tt} d t^2 + g_{rr} d r^2 +
{g}_{\theta \theta} d \theta^2 + {g}_{\theta \theta} \sin^2{\theta} d \phi^2
+ {g}_{t \phi} d t d \phi ,
\end{equation}

\noindent
with  metric components

\begin{eqnarray}
\label{htcomp}
{g}_{tt} & = & - \left(1 - 2 U \right) [1 + 2 K_1 P_2(\cos{\theta})]
- 2 \frac{J^2}{r^4} (2 \cos^2{\theta} - 1) , \nonumber \\
g_{t \phi} & = & - 2 \frac{J}{r} \sin^2{\theta} , \\
g_{rr} & = & \frac{1}{1 - 2 U}
\left[1 - 2 K_2 P_2 (\cos{\theta}) - \frac{2}{1 - 2 U} \frac{J^2}{r^4} \right] ,
\nonumber \\
g_{\theta \theta} & = & r^2 [1 - 2 K_3 P_2 (\cos{\theta})] , \nonumber \\
g_{\phi \phi} & = & g_{\theta \theta} \sin^2{\theta} , \nonumber
\end{eqnarray}

\noindent
where

\begin{eqnarray}
\label{funcs}
K_1 & = & \frac{J^2}{m r^3} (1 + U)
+ \frac{5}{8} \left(\frac{q}{m^3} - \frac{J^2}{m^4} \right)
Q^{2}_{2} \left(\frac{r}{m} - 1 \right) ,
\nonumber \\
K_2 & = & K_1 - \frac{6 J^2}{r^4} , \nonumber \\
K_3 & = & \left(K_1 + \frac{J^2}{r^4} \right)
+ \frac{5}{4} \left(\frac{q}{m^3} - \frac{J^2}{m^4} \right)
\frac{U}{\sqrt{1 - 2 U}} Q^{1}_{2} \left(\frac{r}{m} - 1 \right) , \nonumber
\end{eqnarray}

$$ m = \frac{G M}{c^2} , \quad J = \frac{G S}{c^3} ,
\quad q = \frac{G Q}{c^2} , $$

$$ U = \frac{m}{r} \qquad {\rm and} \qquad
P_2(\cos{\theta}) = \frac{1}{2} [3 \cos^2{\theta} - 1] . $$

\noindent
The functions $ Q^{1,2}_{2} $  are
associated Legendre polynomials of the second kind

$$ Q^{1}_{2} = \sqrt{x^2 - 1} \left(\frac{3}{2} x
\ln{\left(\frac{1 + x}{1 - x} \right)}
- \frac{(3 x^2 - 2)}{(x^2 - 1)} \right) , $$

$$ Q^{2}_{2} = ({x^2 - 1}) \left(\frac{3}{2}
\ln{\left(\frac{1 + x}{1 - x} \right)}
- \frac{(3 x^3 - 5 x)}{(x^2 - 1)^2} \right) . $$

\section{The Post-linear Hartle-Thorne with $ Q^2 $ term}

\noindent
Neglecting $ m^3, \, J^2 $-terms and changing $ q \rightarrow - q $ 
in the HT-metric one obtains

\begin{eqnarray}
\label{htexpand}
g_{tt} & = & - \left(1 - 2 \frac{m}{r} - 2 \frac{q}{r^3} P_2
- 2 \frac{m q}{r^4} P_2 \right) \nonumber \\
g_{t \phi} & = & - 2 \frac{J}{r} \sin^2{\theta} , \\
g_{rr} & = & 1 + 2 \frac{m}{r} + 2 \frac{q}{r^3} P_2
+ 4 \frac{m^2}{r^2} + 10 \frac{m q}{r^4} P_2 \\
g_{\theta \theta} & = & r^2 \left(1 + 2 \frac{q}{r^3} P_2 + 5 \frac{m q}{r^4} P_2
\right) \nonumber \\
g_{\phi \phi} & = & r^2 \sin^2{\theta} \left(1 + 2 \frac{q}{r^3} P_2
+ 5 \frac{m q}{r^4} P_2 \right) . \nonumber
\end{eqnarray}

\noindent
We then added $ q^2 $-terms and checked that the corresponding metric

\begin{eqnarray}
\label{htq2}
g_{tt} & = & - \left(1 - 2 \frac{m}{r} - 2 \frac{q}{r^3} P_2
- 2 \frac{m q}{r^4} P_2 + 2 \frac{q^2}{r^6} P_2^2 \right) \nonumber \\
g_{t \phi} & = & - 2 \frac{J}{r} \sin^2{\theta} , \\
g_{rr} & = & 1 + 2 \frac{m}{r} + 2 \frac{q}{r^3} P_2
+ 4 \frac{m^2}{r^2} + 10 \frac{m q}{r^4} P_2
- \frac{1}{12} \frac{q^2}{r^6} [- 8 P_2^2 + 16 P_2 - 77] \\
g_{\theta \theta} & = & r^2 \left(1 + 2 \frac{q}{r^3} P_2 + 5 \frac{m q}{r^4} P_2
- \frac{1}{36} \frac{q^2}{r^6} [- 44 P_2^2 - 8 P_2 + 43] \right) \nonumber \\
g_{\phi \phi} & = & r^2 \sin^2{\theta} \left(1 + 2 \frac{q}{r^3} P_2
+ 5 \frac{m q}{r^4} P_2
- \frac{1}{36} \frac{q^2}{r^6} [- 44 P_2^2 - 8 P_2 + 43] \right) . \nonumber
\end{eqnarray}

\noindent
is solution of the EFE by means of a REDUCE program \cite{Hearn}.
At this point, it is possible to add rotation into this metric to first order
without problems.

\section{Transformation to Harmonic Coordinates}

\noindent
Harmonic coordinates $ (c T, \, X, \, Y, \, Z) $ are especially useful, because
the form of the metric tensor in these coordinates has a special structure
where the Thorne-moments can be read off directly also for the non-stationary
case \cite{HPM,Thorne}. The harmonic coordinate condition reads

\begin{eqnarray}
\label{hcoord}
\Box X^{\mu} = g^{\alpha \beta} \frac{\partial^2 X^{\mu}}
{\partial x^{\alpha} \partial x^{\beta}}
- g^{\alpha \beta} \Gamma^{\nu}_{\alpha \beta}
\frac{\partial X^{\mu}}{\partial x^{\nu}} = 0 .
\end{eqnarray}

\noindent
The solution of this equation for the HT metric including $ q^2 $-terms reads

\begin{eqnarray}
\label{hcoord2}
T & = & t \nonumber \\
X & = & f \sin{\theta} \cos{\phi} = R \sin{\vartheta} \cos{\phi} \nonumber \\
Y & = & f \sin{\theta} \sin{\phi} = R \sin{\vartheta} \sin{\phi} \\
Z & = & h \cos{\theta} = R \cos{\vartheta} , \nonumber
\end{eqnarray}

\noindent
where

$$ f = \left[r - m + \frac{1}{2} \frac{m q}{r^3} \cos^2{\theta}
+ \frac{1}{72} \frac{q^2}{r^5} (32 P_2^2 - 4 P_2 - 55) \right] $$

$$ h = \left[r - m - \frac{1}{2} \frac{m q}{r^3} \sin^2{\theta}
+ \frac{1}{72} \frac{q^2}{r^5} (32 P_2^2 - 16 P_2 - 43) \right] . $$

\begin{eqnarray}
\label{radius}
R^2 & = & X^2 + Y^2 + Z^2 \nonumber \\
& \simeq & [r - m]^2 + \frac{1}{12} \frac{q^2}{r^4} [8 P_2^2 - 17] \nonumber \\
r & \simeq & R + m - \frac{1}{24} \frac{q^2}{R^5} [8 P_2^2 - 17]
\end{eqnarray}

\begin{eqnarray}
\label{angle}
\tan{\vartheta} & = & \frac{\sqrt{X^2 + Y^2}}{Z} = \frac{f}{h} \tan{\theta}
\nonumber \\
& \simeq & \left[1 + \frac{1}{2} \frac{m q}{r^4}
- \frac{1}{4} \frac{q^2}{r^6} \sin^2{\theta} \right] \tan{\theta} \nonumber \\
\theta & \simeq & \vartheta - \frac{1}{2} \left[\frac{m q}{R^4}
- \frac{1}{2} \frac{q^2}{R^6} \sin^2{\vartheta} \right] .
\end{eqnarray}

\noindent
The transformation of the metric  from the $ (r, \, \theta, \, \phi) $
coordinates to such harmonic coordinates $ (R, \, \vartheta, \, \phi) $,
was performed by means of the differentials 1-forms ($dX, \, dY, \, dZ$).
From these 1-forms, it is solved for the other 1-forms
$ (d r, \, d \theta, \, d \phi) $

\begin{eqnarray}
\label{dr}
d {r} & = & \left(1 - \frac{1}{r} (\alpha1 - \alpha2) \cos^2{\theta}
- \frac{\partial \alpha_1}{\partial r} \sin^2{\theta}
- \frac{\partial \alpha_2}{\partial r} \cos^2{\theta} \right) \nonumber \\
& \times & (\sin{\theta} (\cos{\phi} d X + \sin{\phi} d Y) + \cos{\theta} d Z)
\nonumber \\
& + &
\frac{1}{r} \left(\frac{\partial \alpha_1}{\partial \theta} \sin^2{\theta}
+ \frac{\partial \alpha_1}{\partial \theta} \cos^2{\theta} \right)
(\sin{\theta} d Z - \cos{\theta} (\cos{\phi} d X + \sin{\phi} d Y)) \nonumber \\
& + & \frac{1}{r} (\alpha_1 - \alpha_2) \cos{\theta} d Z
\end{eqnarray}

\begin{eqnarray}
\label{dtheta}
{r} d {\theta} & = & \left(1 + U + U^2
- \frac{1}{r} (\alpha_1 \cos^2{\theta} + \alpha_2 \sin^2{\theta})
- \left(\frac{\partial \alpha_1}{\partial r}
- \frac{\partial \alpha_1}{\partial r} \right)\sin^2{\theta} \right)
\nonumber \\
& \times &(\cos{\theta}(\cos{\phi} d X + \sin{\phi} d Y) - \sin{\theta} d Z)
\nonumber \\
& - & \frac{1}{r} \sin{\theta} \left(\frac{\partial \alpha_1}{\partial \theta}
- \frac{\partial \alpha_2}{\partial \theta} \right)
(\cos^2{\theta}(\cos{\phi} d X + \sin{\phi} d Y) - \cos{\theta} \sin{\theta} dz)
\nonumber \\
& - & \left(\frac{\partial \alpha_1}{\partial r}
- \frac{\partial \alpha_1}{\partial r} \right) \sin{\theta} d Z \\
\label{dphi}
{r} {\sin{\theta}} d {\phi} & = & \left(1 + U + U^2 - \frac{\alpha_1}{r} \right)
(\cos{\phi} d Y - \sin{\phi} d X)
\end{eqnarray}

\noindent
where

\begin{eqnarray}
\alpha_1 & = & \frac{1}{2} \frac{m q}{r^3} \cos^2{\theta}
+ \frac{1}{72} \frac{q^2}{r^5} (32 P_2^2 - 4 P_2 - 55) \nonumber \\
\alpha_2 & = & - \frac{1}{2} \frac{m q}{r^3} \sin^2{\theta}
+ \frac{1}{72} \frac{q^2}{r^5} (32 P_2^2 - 16 P_2 - 43) . \nonumber
\end{eqnarray}


\noindent
Substituting (\ref{dr}), (\ref{dtheta}), and (\ref{dphi}) into the metric
with metric components (\ref{htq2}), the metric in Cartesian coordinates to 
post-linear order takes the form

\begin{eqnarray}
\label{dscart}
d {s}^2 & = & g_{tt} d {t}^2 + 2 g_{t i} d {t} d X^{i} + g_{ij} d X^{i} d X^{j} ,
\end{eqnarray}

\noindent
where


\begin{eqnarray}
\label{htinhc}
g_{tt} & = & - 1 + 2 \frac{w}{c^2} - 2 \frac{w^2}{c^4} \nonumber \\
g_{t X} & = & 2 \frac{J}{R^3} Y , \\
g_{t Y} & = & - 2 \frac{J}{R^3} X , \\
g_{XX} & = & 1 + 2 \frac{w}{c^2} + 2 \frac{w^2}{c^4}
+ \left(\frac{X^2}{R^2} - 1 \right) \frac{m^2}{R^2} \nonumber \\
& + & \frac{1}{2} \frac{m q}{R^4} \left(1 - \frac{X^2}{R^2} - 5 \frac{Z^2}{R^2}
+ 15 \frac{X^2 Z^2}{R^4} \right) \nonumber \\
& + & \frac{q^2}{4 R^6} \left(- 1 + 3 \frac{X^2}{R^2} + 12 \frac{Z^2}{R^2}
- 54 \frac{X^2 Z^2}{R^4} - 15 \frac{Z^4}{R^4}
+ 75 \frac{X^2 Z^4}{R^6} \right) \nonumber \\
g_{XY} & = & m^2 \frac{X Y}{R^4}
+ \frac{1}{2} {m q} \frac{X Y}{R^6} \left(- 1 + 15 \frac{Z^2}{R^2} \right)
+ \frac{3}{4} q^2 \frac{X Y}{R^8}
\left(1 - 18 \frac{Z^2}{R^2} + 25 \frac{Z^4}{R^4} \right) \nonumber \\
g_{XZ} & = & m^2 \frac{X Z}{R^4}
+ \frac{1}{2} {m q} \frac{X Z}{R^6} \left(- 7 + 15 \frac{Z^2}{R^2} \right)
+ \frac{3}{4} q^2 \frac{X Z}{R^8}
\left(5 - 26 \frac{Z^2}{R^2} + 25 \frac{Z^4}{R^4} \right) \nonumber \\
g_{YY} & = & 1 + 2 \frac{w}{c^2} + 2 \frac{w^2}{c^4}
+ \left(\frac{Y^2}{R^2} - 1 \right) \frac{m^2}{R^2} \nonumber \\
& + & \frac{1}{2} \frac{m q}{R^4} \left(1 - \frac{Y^2}{R^2} - 5 \frac{Z^2}{R^2}
+ 15 \frac{Y^2 Z^2}{R^4} \right) \\
& + & \frac{q^2}{4 R^6} \left(- 1 + 3 \frac{Y^2}{R^2} + 12 \frac{Z^2}{R^2}
- 54 \frac{Y^2 Z^2}{R^4} - 15 \frac{Z^4}{R^4}
+ 75 \frac{Y^2 Z^4}{R^6} \right) \nonumber \\
g_{YZ} & = & m^2 \frac{Y Z}{R^4}
+ \frac{1}{2} {m q} \frac{Y Z}{R^6} \left(- 7 + 15 \frac{Z^2}{R^2} \right)
+ \frac{3}{4} q^2 \frac{Y Z}{R^8}
\left(5 - 26 \frac{Z^2}{R^2} + 25 \frac{Z^4}{R^4} \right) \nonumber \\
g_{ZZ} & = & 1 + 2 \frac{w}{c^2} + 2 \frac{w^2}{c^4}
+ \left(\frac{Z^2}{R^2} - 1 \right) \frac{m^2}{R^2} \nonumber \\
& + & \frac{3}{2} \frac{m q}{R^4}
\left(1 - 6 \frac{Z^2}{R^2} + 5 \frac{Z^4}{R^4} \right) \nonumber \\
& + & \frac{3}{4} \frac{q^2}{R^6} \left(- 1 + 15 \frac{Z^2}{R^2}
- 39 \frac{Z^4}{R^4} + 25 \frac{Z^6}{R^6} \right) , \nonumber
\end{eqnarray}

\noindent
where

\begin{equation}
\label{w}
w = \frac{G M}{R} + \frac{G Q}{R^3} P_2 (\cos{\vartheta}) ,
\end{equation}

\noindent
equation (\ref{radius}) was used, and

$$ P_2 (\cos{\vartheta}) = \frac{1}{2} (3 \cos^2{\vartheta} - 1)
= \frac{1}{2} \left(3 \frac{Z^2}{R^2} - 1\right) . $$


\section{Comparison with known results}

\noindent
Our metric (\ref{htinhc}) can directly be compared with the  one
derived in Blanchet \cite{Blanchet1,Blanchet2} that was derived within
the MPM-formalism which works with a tensor field

\begin{equation}
h_{\alpha \beta} = \mathfrak{g}_{\alpha \beta} - \eta_{\alpha \beta} ,
\end{equation}

\noindent
where

$$ \mathfrak{g}_{\alpha \beta} = \sqrt{- g} \, g_{\alpha \beta}
\qquad {\rm and} \qquad g = {\rm det} \, g_{\alpha \beta} $$

\noindent
and uses the Landau-Lifshitz form of the field equations in
harmonic gauge (see e.g., \cite{Blanchet2} for more details).

\noindent
The post-linear metric components under this harmonic coordinates
can be written in the following form

\begin{eqnarray}
\label{blancheteq1}
g_{tt} & = & - \left(1 - \frac{2}{c^2} \, w + \frac{2}{c^4} w^2 \right)
\nonumber \\
g_{ti} & = & g_{it} = - \frac{4}{c^3} w_i \\
g_{ij} & = & \left(1 + \frac{2}{c^2} \, w + \frac{2}{c^4} w^2 \right)
- \frac{4}{c^4} h_{ij} , \nonumber
\end{eqnarray}

\noindent
The potential $ w $ is defined as in (\ref{w}), and

$$ w_{i} = \frac{2 G}{c^2 r^2} {\epsilon}_{i j k} S_{j} n_{k} , $$

\noindent
where $ S_{i} $ is the total angular momentum of the object, and 
$ n_i \equiv x^i/r $ with $ r^2 = x^2 + y^2 + z^2 $. In our case, 
we only have the $ z $ component, {\it i. e.} $ S_z = m a/c $.
  
\noindent
The mass-quadrupole and quadrupole-quadrupole metric components as
obtained by Blanchet \cite{Blanchet1,Blanchet2} take the form

\begin{eqnarray}
\label{blancheteq2}
h^{MM_{ab}}_{00} & = & - \frac{21 M}{r^4} n_{ab} M_{ab} \nonumber \\
h^{MM_{ab}}_{ij} & = & \frac{M}{r^4} \left(- \frac{15}{2} n_{ijab} M_{ab}
- \frac{1}{2} \delta_{ij} n_{ab} M_{ab} + 6 n_{a(i} M_{j)a} - M_{ij}\right)
\nonumber \\
h^{M_{ab}M_{ab}}_{00} & = & \frac{1}{r^6} \left(a^6_0 {\hat{n}}_{abcd} M_{ab} M_{cd}
+ b^6_0 {\hat{n}}_{ab} M_{ac} M_{bc} + c^6_0 M_{ab} M_{ab} \right) \\
h^{M_{ab}M_{ab}}_{ij} & = & \frac{1}{r^6} \left(
p^6_0 {\hat{n}}_{ijabcd} M_{ab} M_{cd} + q^6_0 {\hat{n}}_{ijab} M_{ac} M_{bc}
+ r^6_0 \delta_{ij} {\hat{n}}_{abcd} M_{ab} M_{cd} \right. \nonumber \\
& + & \left. s^6_0 {\hat{n}}_{ij} M_{ab} M_{ab}
+ t^6_0 \delta_{ij} {\hat{n}}_{ab} M_{ac} M_{bc} + u^6_0 \delta_{ij} M_{ab} M_{ab}
+ v^6_0 {\hat{n}}_{abc(i} M_{j)a} M_{bc} \right. \nonumber \\
& + & \left. w^6_0 {\hat{n}}_{a(i} M_{j)b} M_{ab}
+ x^6_0 {\hat{n}}_{ab} M_{ij} M_{ab} + y^6_0 {\hat{n}}_{ab} M_{a(i} M_{j)b}
+ z^6_0 M_{a(i} M_{i)a} \right) , \nonumber
\end{eqnarray}

\noindent
where

$$ a^6_0 = - \frac{63}{4} , \qquad  b^6_0 = - 9 ,
\qquad c^6_0 = - \frac{21}{10} , $$

$$ p^6_0 = - \frac{75}{4} , \qquad  q^6_0 = \frac{90}{11} ,
\qquad r^6_0 = - \frac{9}{44} , $$

$$ s^6_0 = \frac{25}{84} , \qquad  t^6_0 = - \frac{29}{42} ,
\qquad u^6_0 = - \frac{11}{70} , $$

$$ v^6_0 = - \frac{18}{11} , \qquad  w^6_0 = \frac{5}{21} ,
\qquad x^6_0 = - \frac{10}{21} , $$

$$ y^6_0 = \frac{23}{42} , \qquad z^6_0 = \frac{6}{35} , $$

\noindent
and $ {\hat{n}}_{i_1 \dots i_l} $ are the  symmetric and trace-free
parts (e.g., \cite{Poisson,Thorne}) of the Cartesian tensor

\begin{equation}
\label{ns}
n_{i_1 \dots i_l} \equiv n_{i_1} \cdots n_{i_l} .
\end{equation}

\noindent
For our axially symmetric body the quadrupole-moment $ Q $ appears in
the Cartesian quadrupole mass-tensor in the form

\begin{equation}
\label{quadro}
M_{ij}  =  - \frac{Q}{3} (\delta_{ij} - 3 \delta_{i3} \delta_{j3} ) .
\end{equation}

\noindent
Substituting these $ M_{ab} $ components into (\ref{blancheteq2}) and
the resulting $ h_{\alpha \beta} $ into (\ref{blancheteq1}), we get
the metric components (\ref{htinhc}).

\section{Transformation from Erez-Rosen to Hartle-Thorne}

\noindent
The Erez-Rosen (ER) metric represents an static exact solution of EFE with
axial symmetry and a quadrupole moment
\cite{Carmeli,Doroshkevich,Winicour,Young,Zeldovich}.
Keeping only $ q^2 $ terms the ER-metric in spherical coordinates
$ (ct, \, r, \, \theta, \, \phi) $ reads

\begin{eqnarray}
\label{er}
g_{tt} & = & - \left(1 - 2 U - \frac{4}{15} q U^3 P_2 - \frac{4}{15} q U^4 P_2
+ \frac{8}{225} q^2 U^6 P_2^2 \right) \nonumber
\end{eqnarray}

\begin{eqnarray}
g_{rr} & = & 1 + 2 U + 4 U^2 + \frac{4}{15} q U^3 P_2
+ \frac{4}{45} q U^4 (5 P_2^2 + 11 P_2 - 1) \nonumber \\
& + & \frac{8}{2025} q^2 U^6 (25 P_2^3 - 12 P_2^2 - 6 P_2 + 2) \\
g_{\theta \theta} & = & r^2 \left(1 + \frac{4}{15} q U^3 P_2
+ \frac{4}{45} q U^4 (5 P_2^2 + 5 P_2 - 1) \right. \nonumber \\
& + & \left. \frac{8}{2025} q^2 U^6 (25 P_2^3 - 12 P_2^2 - 6 P_2 + 2)
\right) \nonumber \\
g_{\phi \phi} & = & r^2 \sin^2{\theta} \left(1 + \frac{4}{15} q U^3 P_2
+ \frac{4}{5} q U^4 P_2 + \frac{8}{225} q^2 U^6 P_2^2 \right) , \nonumber
\end{eqnarray}

\noindent
where

$$ q = \frac{Q}{m^3} \qquad {\rm and} \qquad U = \frac{m}{r} . $$

\noindent
The following transformation converts the ER truncated metric into the static 
HT metric at the same level of approximation

\begin{eqnarray}
r & = & R \left(1 + \frac{m q}{R^4} f_1 + \frac{q^2}{R^6} f_2 \right)
\nonumber \\
\theta & = & \Theta + \frac{m q}{R^4} g_1 + \frac{q^2}{R^6} g_2
\end{eqnarray}

\noindent
where

\begin{eqnarray}
f_1 & = &  \frac{1}{135} (10 P^2_2 - 8 P_2 - 2) \nonumber \\
f_2 & = & \frac{1}{4050} (40 P^3_2 - 24 P^2_2 - 43)
\nonumber \\
g_1 & = & - \frac{1}{45} (5 P_2 - 2) \cos{\Theta} \sin{\Theta} \nonumber \\
g_2 & = & \frac{P_2}{2025} (12 - 30 P_2) \cos{\Theta} \sin{\Theta}
\nonumber
\end{eqnarray}

\noindent
with $ P_2 = P_2(\cos{\Theta}) $.

\noindent
The transformed metric components are given by

\begin{eqnarray}
\label{erht}
g_{tt} & = & - \left(1 - 2 \, {\cal U} - \frac{12}{45} q \, {\cal U}^3 P_2
- \frac{12}{45} q \, {\cal U}^4 P_2 + \frac{8}{225} q^2 \, {\cal U}^6 P_2^2
\right) \\
g_{R \Theta} & \simeq & 0 \nonumber \\
g_{RR} & = & 1 + 2 \, {\cal U} + 4 \, {\cal U}^2
+ \frac{12}{45} q \, {\cal U}^3 P_2
+ \frac{4}{3} q \, {\cal U}^4 P_2 \nonumber \\
& + & \frac{1}{315} q^2 \, {\cal U}^6
\left(\frac{168}{45} P_2^2 - \frac{336}{45} P_2 + \frac{1617}{45} \right)
\nonumber \\
g_{\Theta \Theta} & = & R^2 \left(1 + \frac{12}{45} q \, {\cal U}^3 P_2
+ \frac{30}{45} q \, {\cal U}^4 P_2 \right. \nonumber \\
& + & \left. \frac{1}{315} q^2 \, {\cal U}^6 \left(
\frac{308}{45} P_2^2 + \frac{56}{45} P_2 - \frac{301}{45} \right) \right)
\nonumber \\
g_{\phi \phi} & = & R^2 \sin^2{\Theta}
\left(1 + \frac{12}{45} q \, {\cal U}^3 P_2 + \frac{30}{45} q \, {\cal U}^4 P_2
\right. \nonumber \\
& + & \left. \frac{1}{315} q^2 \, {\cal U}^6 \left(
\frac{308}{45} P_2^2 + \frac{56}{45} P_2 - \frac{301}{45} \right)
\right) , \nonumber
\end{eqnarray}

\noindent
where $ {\cal U} = {m}/ {R} $.

\noindent
Changing $ q \rightarrow 15q/2 $, one obtains the static HT metric at this
level of approximation.

\section{Conclusions}

\noindent
We expanded the HT-metric and kept only linear terms in the rotation parameter 
and quadratic terms in the mass parameter. Then we included second order terms 
in the quadrupole parameter by solving the EFE in vacuum perturbatively. 
We then shows that this form of the metric agrees with a corresponding metric 
that was derived within the MPM-formalism by Blanchet at the same order of 
approximation.

\noindent
A transformation linking our static HT solution with the ER metric expanded 
in Taylor series was also found. This provides a validation of all these 
metrics. The quadrupole moment is an important feature which is included as 
a physical parameter in all these solutions.

\noindent
These spacetimes can be used to represent realistic solutions of
self-gravitating (axially symmetric) mass distribution of perfect fluid. This
is because the HT solution can be smoothly matched with interior perfect
fluid solution with physically reasonable properties.

\end{document}